\documentclass[linenumbers]{aastex631}

\shorttitle{Brightest galaxies in Planck clusters}
\shortauthors{Smith J. C., et al.}

\graphicspath{{./}{figures/}}

\begin{document}

\title{Towards discovery of gravitationally lensed explosive transients: the brightest galaxies in massive galaxy clusters from Planck-SZ2}
\author[0000-0002-7276-5905]{Joshua C. Smith}
\author[0000-0002-4429-3429]{Dan Ryczanowski}
\author[0000-0002-0427-5373]{Matteo Bianconi}
\author{Denisa Cristescu}
\author{Sivani Harisankar}
\author{Saskia Hawkins}
\author{Megan L.\ James}
\author{Evan J.\ Ridley}
\author{Simon Wooding}
\author[0000-0003-4494-8277]{Graham P. Smith}
\affil{School of Physics and Astronomy, University of Birmingham, Edgbaston, B15 2TT, U.K.}

\begin{abstract}
\noindent We combine the Planck-SZ2 galaxy cluster catalogue with near-infrared photometry of galaxies from the VISTA Hemisphere Survey to identify candidate brightest cluster galaxies (BCGs) in 306 massive clusters in the Southern skies at redshifts of $z>0.1$. We find that 91\% of these clusters have at least one candidate BCG within the 95\% confidence interval on the cluster centers quoted by the Planck collaboration, providing reassurance that our analyses are statistically compatible, and find 92\% to be reasonable candidates following a manual inspection. We make our catalog publicly available to assist colleagues interested in multi-wavelength studies of cluster cores, and the search for gravitationally lensed explosive transients in upcoming surveys including the Legacy Survey of Space and Time by the Vera C. Rubin Observatory. 
\end{abstract}

\section{Introduction} \label{sec:intro}

The Vera C. Rubin Observatory's (Rubin's) upcoming Legacy Survey of Space and Time (LSST) will detect and rapidly release millions of alerts per night, 
and is expected to include, for example, hundreds of gravitationally lensed supernovae per year \citep{2019ApJS..243....6G}.
Recent work has also shown that highly magnified gravitationally lensed kilonova counterparts to gravitationally lensed binary neutron star mergers will be detectable with Rubin \citep{Smith2023}. 

The optical depth to strong gravitational lensing spans the dark matter halo mass function from individual massive early type galaxies through to the most massive galaxy clusters \citep{2020MNRAS.495.3727R}. An all-sky list of clusters with accurate coordinates is therefore essential in the search for lensed explosive transients of all flavours 
\citep{2022arXiv220412984R}, especially in the context of lensed transients that are magnified by a factor $\mu\gtrsim10$ \citep[e.g.][]{Smith2023}. In this Research Note we describe work done primarily by second year undergraduate students at the University of Birmingham to refine the centers of massive clusters in the Planck 2nd Sunyaev-Zeldovich Source Catalog \citep[PSZ2,][]{2016A&A...594A..27P}. This is as an input to ongoing and future searches for lensed explosive transients, and may be of interest for other research involving BCGs/cluster cores.

We assume a flat cosmology with $H_0=67.9\,\rm km\,s^{-1}\,Mpc^{-1}$, $\Omega_{\rm M}=0.3065$ \citep{Planck15cos}. All magnitudes are stated in the Vega system.

\section{Data and methods}
\label{sec:data}

PSZ2 is 
an all-sky catalog of clusters detected by the \emph{Planck} satellite via the Sunyaev-Zeldovich effect. The typical positional uncertainty on these clusters is a few arcminutes, i.e.\ much larger than the typical Einstein radius of a cluster, $\theta_{\rm E}\lesssim20$\,arcsec. Moreover, the strong lensing region of clusters is centered on the brightest few galaxies within a cluster core, which may be offset from signals emanating from the intracluster medium \citep[e.g.][]{Smith2005}. We therefore aim to refine the PSZ2 cluster coordinates by finding candidate BCGs in these clusters. We do this using $J$- and $K_{\rm S}$-band (hereafter $K$-band) photometry of galaxies from the VISTA Hemisphere Survey \citep[VHS,][]{2013Msngr.154...35M}.

A total of $306$ PSZ2 clusters reside within the VHS footprint with redshifts of $z\geq0.1$. This lower cut on redshift was motivated by the strongly declining optical depth to strong lensing at very low redshifts. We estimated the projected angular radius, $\theta_{500}$, that corresponds to the physical radius $r_{500}$ for each of these clusters from the mass and redshift information provided in the PSZ2 catalog. These angular radii were typically twice the quoted 95\% confidence interval on the cluster center (hereafter, $\theta_{\rm P}$). We then selected extended ($p_{\rm gal}>0.5$) securely detected ($\sigma_K<0.2$) sources from the VHS catalog located within $\theta<\theta_{500}$ of each cluster center. The analysis described below used extinction corrected magnitudes measured in apertures of diameter $2\,\rm arcsec$. 

We use a galaxy evolution model to estimate the $(J-K)$ colours of passively evolving $L^\star$ cluster galaxies as a function of redshift, based on the \cite{2003MNRAS.344.1000B} stellar evolution model and adopting a single burst of star formation at $z=3$ with solar metallicity, and the \cite{2003PASP..115..763C} initial mass function. The model was normalised to an $L^\star$ cluster galaxy at $z=0.25$, i.e.\ $K^\star(z=0.25)=15.5$, based on \cite{2004ApJ...610..745L}. 

\begin{figure}
\includegraphics[width=\textwidth]{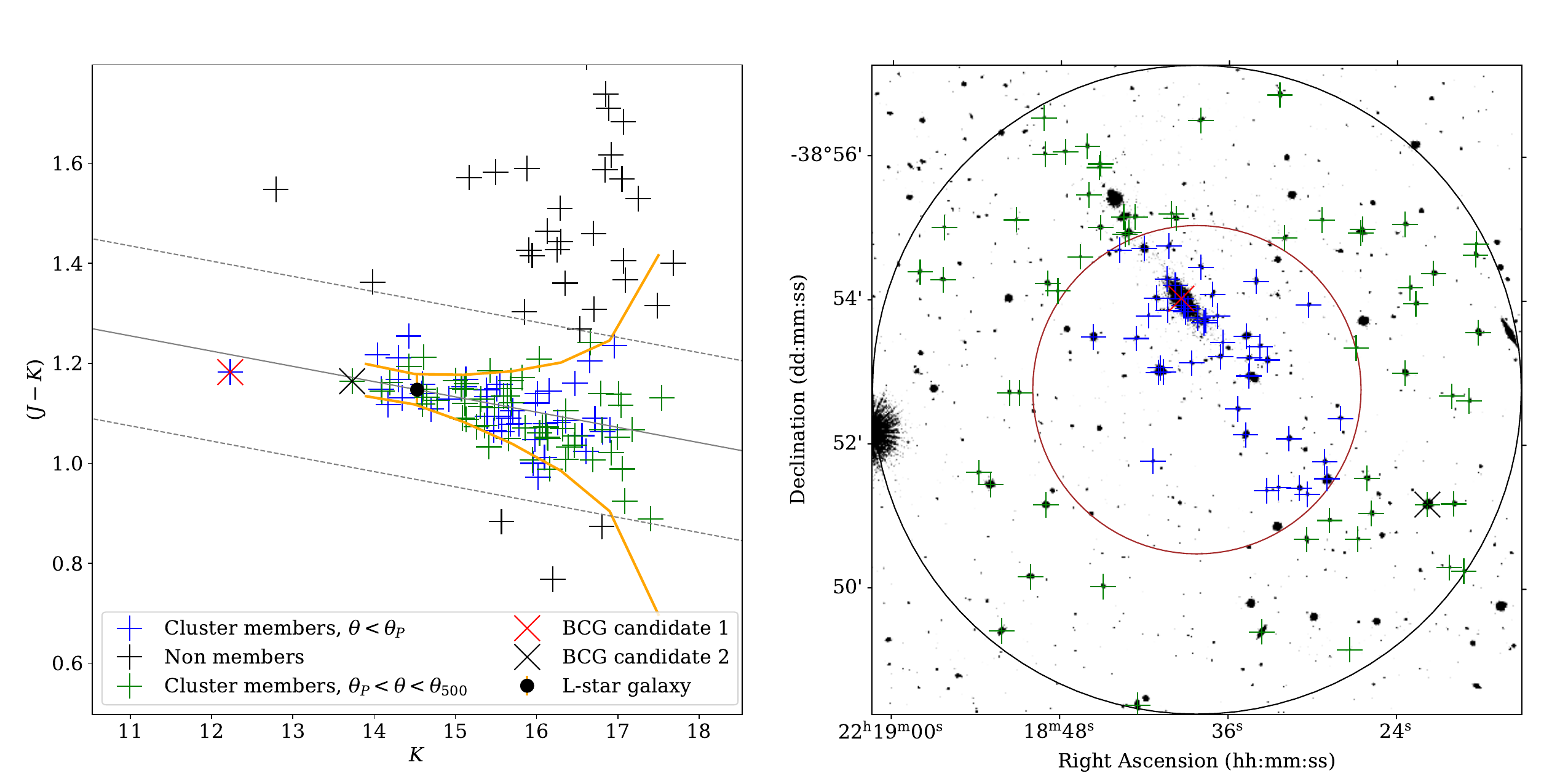}
\caption{{\sc Left} -- Colour-magnitude diagram for an example cluster.
Expected colors of cluster members are bounded by the orange lines, based on the methods described in Section~\ref{sec:data}. {\sc Right} -- The sky locations of the galaxies seen at left are marked on a cut-out image from the VHS survey. The inner and outer circles are centered on the cluster center provided by PSZ2 and mark $\theta_{\rm P}$ and $\theta_{500}$ respectively, where $\theta_{\rm P}$ is the 95\% confidence interval on cluster position quoted by PSZ2.
\label{fig:plot}}
\end{figure}

The two brightest galaxies that satisfy $K^\star-3.5<K<K^\star$ and $|\Delta(J-K)|\le0.2$ were selected as candidate BCGs for each cluster. The bright cut on $K$ removes any contaminants that leaked through the selection criteria applied to the VHS catalogs, and is sufficiently generous to ensure we retain the brightest BCGs \cite[][]{2008MNRAS.384.1502S}. 
$\Delta(J-K)$ is the offset of each source from the color of the ridge line galaxies (often referred to as the red sequence) at the cluster redshift, predicted from our model for $K^\star(z)$ and the slope, $m(z)$, of this ridge line as a function of redshift from \cite{2009MNRAS.394.2098S}.

\section{Results and public data release} 
\label{sec:results}
Following manual inspection of colour-magnitude diagrams and the sky distributions of candidate member galaxies for each cluster, we are confident in the automated identification of BCG candidates for 275 of the 306 cluster sample, as they show convincing evidence of a ridge line of cluster members consistent with the redshift given in the PSZ2 catalog (see Figure~\ref{fig:plot}). A further 6 clusters have convincing ridge lines of cluster members at a $(J-K)$ color that is discrepant with the redshift listed in the PSZ2 catalog. In these cases, we forego the Planck redshift and instead infer a new redshift from the ridge line, using this to determine $L^*$ and select BCG candidates. Therefore, we are confident with 281/306 (92\%) of the BCG detections made via this method. The remaining 25 clusters that do not show a ridge line all have an estimate of $L^*$ that is within $\sim 0.1$ mag or fainter than the VHS 5$\sigma$ limit, highlighting the limitations of this method due to survey depth. 

Of the 281 confident detections, 4 clusters were contaminated by  
interloping foreground stars/galaxies not removed by our selection. In these cases, the contaminants were manually removed.
Finally, in 91\% of cases the brightest galaxy resides within the 95\% confidence interval on the cluster centre's position from PSZ2, providing reassurance that our analysis is statistically compatible with the Planck collaboration's analysis.

We make our candidate BCGs publicly available as a machine readable table listing all 306 clusters and containing the results of our automated identification of two BCG candidates for each cluster, plus a flag that identifies to which of the categories discussed above each cluster belongs. If it is one of the seven clusters with red sequences at discrepant redshifts, we append the proposed alternative redshift used within our analysis. The table and full details of its content is available via the CDS archive.

We thank staff and colleagues at the University of Birmingham's School of Physics and Astronomy for enabling this work to be completed as part of an undergraduate project. DR, MB and GPS acknowledge support from STFC (grant numbers: ST/N021702/1 and ST/S006141/1). GPS acknowledges support from The Royal Society and the Leverhulme Trust. This Research Note is based on observations obtained as part of the VISTA Hemisphere Survey, ESO Progam, 179.A-2010 (PI: McMahon), and on observations obtained with \emph{Planck} (http://www.esa.int/Planck), an ESA science mission with instruments and contributions directly funded by ESA Member States, NASA, and Canada.



\bibliography{sample631}{}
\bibliographystyle{aasjournal}

\end{document}